\documentclass[12pt]{article}
\usepackage{epsfig}
\usepackage{amsfonts}
\begin{document}
\begin{titlepage}
\begin{center}

{\Large Chaotic quantization and the parameters of the standard
model}

\vspace{2.cm} {\bf Christian Beck}

\vspace{2.cm}

School of Mathematical Sciences, Queen Mary, University of London,
Mile End Road, London E1 4NS, UK.

\vspace{5cm}

\end{center}

\abstract{In the chaotic quantization approach one replaces the
Gaussian white noise of the Parisi-Wu approach of
stochastic quantization by a deterministic chaotic process on a
very small scale. We consider suitable coupled chaotic noise
processes as generated by Tchebyscheff maps, and show that the
vacuum energy of these models is minimized for coupling constants
that coincide with running standard model couplings at energy
scales given by the known fermion and boson masses. Chaotic
quantization thus allows to predict fundamental constants of
nature from first principles. At the same time, it provides a
natural framework to understand the dynamical origin of vacuum
energy in our universe.}

\vspace{1.3cm}

\end{titlepage}

\section{Introduction}

The important role of chaos in quantum field theories and string
theories has been emphasized in various recent papers and books
\cite{thooft}--\cite{biro2}. `t Hooft conjectures that the
ultimate theory underlying quantum mechanical behaviour is a
dissipative one exhibiting complex behaviour \cite{thooft}.
Damour, Henneaux, Julia and Nicolai \cite{nico} emphasize that M-theory, the
hypothetical theory of all interactions, is intrinsically chaotic.
Kogan and Polyakov have recently studied chaotic renormalization flow
and Feigenbaum universality in string theory \cite{poly}. An
important way how chaos can enter into quantum field theories is
via the so-called 'chaotic quantization' method \cite{nonlin}.
Here one assumes that the noise used for stochastic quantization
has a dynamical (deterministic chaotic)
origin. It has been recently pointed out \cite{physicad,book}
that this method
allows for some very precise predictions of
standard model parameters. Also, Bir\'{o}, M\"uller
and Matinyan have recently shown
that chaotic classical Yang Mills theories \cite{biro1} can
'quantize themselves', i.e.\ the noise used for stochastic
quantization can be intrinsically generated by the strongly
chaotic behaviour of the classical field equation \cite{biro2}.

In the following sections, we will first review how to extend the
stochastic quantization approach of Parisi and Wu
\cite{stoch1,stoch2} to a chaotic quantization method
\cite{nonlin}. Then we summarize how this method can yield
predictions on the fundamental constants of nature, such as
coupling constants, masses, and mixing angles of the standard
model, using a simple principle, the minimization of vacuum energy
\cite{physicad,book}. Indeed, one of the main features of the
chaotic quantization approach is that it naturally produces a
non-vanishing expectation of vacuum energy, due to the potentials
of the underlying deterministic chaotic theory on the smallest
scales. This vacuum energy may well stand in relation to the dark
energy that is currently observed in our universe \cite{dark}. In
this paper we just review the main ideas and results, much more
details can be found in \cite{book}.

\section{Chaotic quantization}

Let us first recall the stochastic quantization method, then we
generalize it to chaotic quantization. A field theory is usually
determined by some action functional $S[\phi]$. The field $\phi$
is a function of the space-time coordinates and may, in general,
have many components. The classical field equation can be written
as
\begin{equation}
\frac{\delta S}{\delta \phi} =0,
\end{equation}
meaning that the action has an extremum.

In the Parisi-Wu appraoch of stochastic quantization one proceeds
from the classical field equation to a quantized theory by means
of the following Langevin equation:
\begin{equation}
\frac{\partial}{\partial t}\phi (x,t)=-\frac{\delta S}{\delta
\phi}(x,t) + L(x,t)                                      \label{1}
\end{equation}
Here $x=(x^1,x^2,x^3,x^4)=x^\mu$ is a point in Euclidean
space-time, $t$ denotes a fictitious time variable (different from
the physical time $x^4$), and $L(x,t)$ denotes spatio-temporal
Gaussian white noise, $\delta$-correlated in both space-time
$x$ and fictitious time $t$.

The fictitious time $t$ is just introduced as an artificial fifth
coordinate. It is different from the physical time. What is of
physical relevance is the stationary solution of the Langevin
equation in the limit $t \to \infty$. It is the quantized field, a
stochastic process. All quantum mechanical expectations of the
field $\phi (x)$ can be calculated as expectations with respect to
the realizations of the Langevin process in the limit $t\to
\infty$.

The action $S$ of the entire standard model can be 2nd-quantized
in this way (at least in principle). For each standard model
field, there is a corresponding noise field. One may then ask:
Where do these rapidly fluctuating noise fields ultimately come
from? Could they have dynamical origin? The idea of chaotic
quantization is that the noise fields used for second quantization
are not truely random but generated by a rapidly fluctuating
deterministic chaotic process. One can, for example, generate the
noise variables at each space-time point by a chaotic map $T$. If
this map has the so-called $\varphi$-mixing property \cite{billing}, then it can
be rigorously proved that rescaled sums of iterates generate the
Wiener process (= Brownian motion) on large scales, regarding the
initial value as a random variable. In other words, the fast
chaotic dynamics looks locally like Gaussian white noise if seen from a
larger scale. Hence, on large scales ordinary quantum field
theoretical behaviour is generated if chaotic 'noise' is used for
quantization. Only on small scales (the Planck scale or below)
there is much more complex behaviour and nontrivial correlations.

A simple model is to generate the noise by Tchebyscheff maps.
One can actually show that Tchebyscheff maps of order $N$
\begin{equation}
\Phi_{n+1}=T_N(\Phi_n),\;\;\;\;\;\Phi_0\in [-1,1],
\end{equation}
are $\varphi$-mixing \cite{billing}. 
In nonlinear dynamics, the $T_N$ are standard examples
of chaotic maps, just similar as the harmonic oscillator is a
standard example in linear dynamics. One has $T_2(\Phi)=2\Phi^2-1$
and $T_3(\Phi)=4\Phi^3-3\Phi$, generally $T_N(\Phi )=\cos (N
\arccos \Phi )$.
There is
sensitive dependence on initial conditions for $N\geq 2$: Small perturbations in
the initial values will lead to completely different trajectories
in the long-term run. The maps are conjugated to a Bernoulli shift
with an alphabet of $N$ symbols. This means, in suitable
coordinates the iteration process is just like shifting symbols in
a symbol sequence.

Most important for our purposes is the following property: One can
show that the Tchebyscheff maps have least higher-order
correlations among all smooth systems conjugated to a Bernoulli
shift, and are in that sense closest to Gaussian white noise, as
close as possible for a smooth deterministic system
\cite{non91,hilgers}. Any other map has more higher-order
correlations. What does this mean for chaotic quantization? It is
plausible that if nature chooses to generate Gaussian white noise
by something deterministic chaotic on the smallest quantization
scales, it aims for making the small-scale deviations from
ordinary quantum mechanics as small as possible. This
automatically leads to Tchebyscheff maps. A graph theoretical
method for this type of `deterministic noise' has been developed
in \cite{non91,hilgers}.

\section{Coupled chaotic noise fields}

Once we assume that the noise fields used for quantization are
dynamical in origin, it is natural to allow for some coupling
between neighbored noise fields.
In string theory, in a perturbative approach, point particles are
replaced by little extended 1-dimensional objects, strings. Now if
we go to strings in the standard model space, it's
natural to also proceed to 'chaotic strings' in the corresponding
chaotic noise space used for second quantization. This is
illustrated in Fig.~1. Each ordinary string might be 'shadowed' by
a corresponding chaotic noise string used for second quantization
purposes.
\begin{figure}
\hspace{3cm}
\epsfig{file=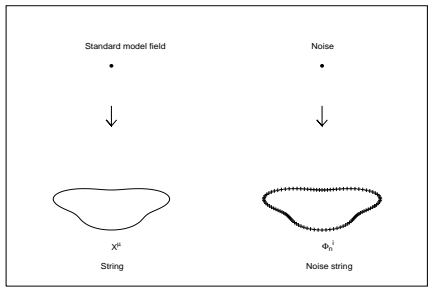, width=6cm, height=4cm} \caption{In
string theory, point particles are replaced by strings. For
symmetry reasons, we may then also replace the
chaotic point noise fields used for 2nd quantization by little
extended noise objects, 'chaotic strings'. }
\end{figure}

Among the many models that can be chosen to generate a coupled
chaotic dynamics on a small scale certain criteria should be
applied to select a particular system. First of all, for vanishing
spatial coupling of the chaotic `noise' one wants to have
strongest possible random behavior with least possible
higher-order correlations, in order to be closest to the Gaussian
limit case (which corresponds to ordinary path integrals on a
large scale). This selects as a local dynamics Tchebyscheff maps
$T_N(x)$ of $N$-th order ($N\geq 2$). Now let us discuss possible
ways of spatially coupling the chaotic noise. Although in
principle all types of coupling forms can be considered,
physically it is most reasonable that the coupling should result
from a Laplacian coupling rather than some other coupling, since
this is the most relevant coupling form in quantum field and
string theories. This leads to coupled map lattices of the
nearest-neighbour coupling form. The resulting coupled map
lattices can then be studied on lattices of arbitrary dimension,
but motivated by the fact that ordinary strings are 1-dimensional
objects we will here consider 1-dimensional structures, although
higher-dimensional chaotic objects (`chaotic branes') can be
studied as well \cite{pla,book}. We end up with coupled
Tchebyscheff maps of the form
\begin{equation}
\Phi_{n+1}^i=(1-a)T_N(\Phi_n^i) + s \frac{a}{2} (T_N^b
(\Phi_n^{i-1}) +T_N^b(\Phi_n^{i+1})), \label{dyni}
\end{equation}
where $i$ is a 1-dimensional lattice coordinate, $a\in [0,1]$ is a
coupling constant, $s=\pm 1$, and $b$ takes on values 0 or 1
($T^0(\Phi)=\Phi, T^1(\Phi)=T(\Phi)$). The chaotic string
dynamics
(\ref{dyni}) is deterministic chaotic, spatially extended, and
strongly nonlinear. The field variable $\Phi_n^i$ is physically
interpreted in terms of rapidly fluctuating virtual momenta in
units of some arbitrary maximum momentum scale.

It is easy to see that for odd $N$ the statistical properties of
the coupled map lattice are independent of the choice of $s$
(since odd Tchebyscheff maps satisfy $T_N(-\Phi )=-T_N(\Phi )$),
whereas for even $N$ the sign of $s$ is relevant and a different
dynamics arises if $s$ is replaced by $-s$. Hence, restricting
ourselves to the `ground states' of our chaotic string
oscillators, i.e. $N=2$ (even maps) and $N=3$ (odd maps), in total
6 different chaotic string theories arise, characterized by
$(N,b,s)=(2,1,+1),(2,0,+1),(2,1,-1),(2,0,-1)$ and
$(N,b)=(3,1),(3,0)$. For easier notation, we have labeled these
chaotic string theories as $2A,2B,2A^-,2B^-,3A,3B$, respectively.
Chaotic strings can also be regarded as discrete versions of
self-interacting scalar fields that are homogeneous in all but one
space-time direction.

\section{Vacuum energy density due to chaotic quantization effects}

Though the chaotic string dynamics is dissipative, one can
formally introduce potentials that generate the discrete time
evolution. For example, the 3rd-order Tchebyscheff dynamics can be written as
\begin{equation}
\Phi_{n+1}-\Phi_n=4 \Phi_n^3-4\Phi_n.
\end{equation}
This equation formally
describes a discrete momentum change (= force) generated by the
self-interacting potential
\begin{equation}
V^{(3)} (\Phi )= \left( -\Phi^4+\frac{3}{2} \Phi^2 \right)
+\frac{1}{2} \Phi^2 +C , \label{sumv3}
\end{equation}
the force being given by $-\frac{\partial V^{(3)}}{\partial
\Phi}$. One can now incorporate symmetry considerations between
$+T_N$ and $-T_N$ \cite{book}, with the result that there are two interesting
observables to look at for chaotic strings, the expectation of the
self energy given by

\begin{eqnarray}
V^{(2)}(a) &=& -\frac{2}{3} \langle \Phi^3 \rangle +\langle \Phi
\rangle \;\;\;\;(N=2)\label{A} \\ V^{(3)}(a) &=&  -\langle \Phi^4 \rangle
+\frac{3}{2} \langle \Phi^2 \rangle \;\;\;\; (N=3),
\end{eqnarray}
and the expectation of the interaction energy given by
\begin{equation}
W(a) =\frac{1}{2} \langle \Phi^{i} \Phi^{i+1} \rangle . \label{B}
\end{equation}
All expectations can be calculated as long-term averages over $n$
and $i$ for random initial conditions. 

Note that in quite a
natural way there is vacuum energy associated with our
deterministic chaotic dynamics, given by the above equations.
Could this vacuum energy have something to do with the dark energy
that makes up most of the energy density (70$\%$) of our universe,
as recently confirmed by various astronomical observations? It
might indeed. The absolute unit of the vacuum energy of our
chaotic noise fields is not fixed in our theory.
But most naturally, if we
quantize a particle of mass $m$ then one would expect that the
corresponding vacuum energy of the corresponding noise
field yields a similar energy contribution, since the potential
$V$ simply generates chaotic fluctuations of the particle
momentum $mc$ (\cite{book}, chapter 5). 
Hence it is most natural to conjecture that the vacuum
energy density generated by chaotic quantization effects has the
same order of magnitude as the mass density of particles in our
universe, since for each particle
there is a corresponding noise field used for
quantization. This could point towards a possible solution of the
'cosmological coincidence' problem \cite{coin}.

The expectations of vacuum energy (\ref{A})--(\ref{B}) depend on the coupling
constant $a$ in a nontrivial way and are like a 'thermodynamic
potential' of vacuum fluctuations. $V(a)$ and $W(a)$ (and their
sum \cite{book}) can be easily numerically determined by iterating the coupled
map and averaging over all $i$ and $n$.

A helpful physical interpretation of the coupled map dynamics is
as follows. Suppose we regard $\Phi_n^i$ to be a fluctuating
virtual momentum component that can be associated with a hypothetical
'noise' particle $i$ at time $n$, all particles
$i$ being ultimately responsible for a dynamical state
underlying dark energy. $n$ can be either interpreted as
fictitious time or as physical time, both interpretations
make sense
\cite{book}. Neighbored
particles $i$ and $i-1$ can exchange momenta due to the Laplacian
coupling of the coupled map lattice. We may actually associate a
fermion-antifermion pair $f_1,\bar{f}_2$ with each cell $i$. In
units of some arbitrary energy scale $p_{max}$, the particle has
momentum $\Phi_n^i$, the antiparticle momentum $-\Phi_n^i$. They
interact with particles in neighbored cells by exchange of a
(hypothetical) gauge boson $B_2$, then they annihilate into
another boson $B_1$ until the next vacuum fluctuation takes place.
This can be (symbolically!) described by the Feynman graph in Fig.~2.
\begin{figure}
\hspace{3cm}
\epsfig{file=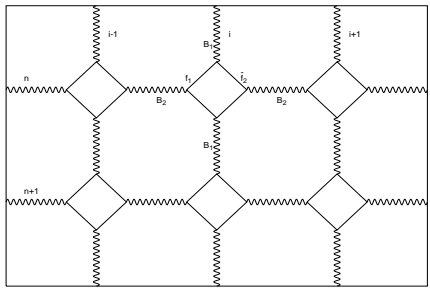, height=4cm, width=6cm} \caption{Interpretation of the coupled
map dynamics in terms of fluctuating momenta exchanged by fermions
$f_1,\bar{f}_2$ and bosons $B_1,B_2$. $\Phi_n^i$ corresponds to the
momentum in the fermion loop.}
\end{figure}
We call this graph a `Feynman web', since it describes an
extended spatio-temporal interaction state of space-time, to which
we have given a standard model-like interpretation. Note that in
this interpretation $a$ is a (hypothetical) standard model
coupling constant, since it describes the strength of momentum
exchange of neighbored particles. At the same time, $a$ can also
be regarded as an inverse metric in the 1-dimensional string
space, since it determines the strength of the Laplacian coupling.

What is now observed numerically for the various chaotic strings
is that the interaction energy $W(a)$ has zeros and the self
energy $V(a)$ has local minima for string couplings $a$ that
numerically coincide with running standard model couplings
$\alpha(E)$, the energy being given by
\begin{equation}
E= \frac{1}{2} N \cdot (m_{B_1}+m_{f_1}+m_{f_2}).
\label{katharina}
\end{equation}
Here $N$ is the index of the chaotic string theory considered, and
$m_{B_1}, m_{f_1}$, $m_{f_2}$ denote the masses of the standard
model particles involved in the Feynman web interpretation. The
surprising observation is that rather than yielding just some
unknown exotic physics, the chaotic string spectrum appears to
reproduce the masses and coupling constants of the known quarks,
leptons and gauge bosons of the standard model with very high
precision. Gravitational and Yukawa couplings are observed as
well. The chaotic dynamics can be used to fix the fundamental
constants of nature by a simple principle, the minimization of
vacuum energy.

\section{Some numerical results}

Let us now present some examples of numerical results (much more
numerical evidence on the validity of eq.~(\ref{katharina}) can be
found in \cite{book}). Fig.~3 shows the interaction energy
$W(a)=\frac{1}{2} \langle \Phi_n^i \Phi_n^{i+1} \rangle$ of the
chaotic $3A$ string in the low-coupling region.
\begin{figure}
\hspace{3cm}
\epsfig{file=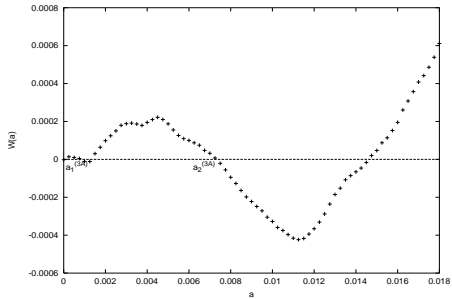, height=4cm, width=6cm}
\caption{Numerically determined interaction energy of the 3A
string.}
\end{figure}
We numerically find two zeros of $W(a)$ with $W'(a)<0$ in the
low-coupling region:
\begin{eqnarray*}
a_1^{(3A)}&=&0.0008164(8)     \\ a_2^{(3A)}&=&0.0073038(17)
\end{eqnarray*}
Remarkably, the zero $a_2^{(3A)}$ appears to coincide with the
running fine structure constant $\alpha_{el} \approx 1/137$,
evaluated at an energy scale given by 3 times the electron mass.
We find the amazing numerical coincidence
\begin{equation}
a_2^{(3A)}=\alpha_{el}(3m_e), \label{ael}
\end{equation}
the energy scale $3m_e$ being in agreement with
eq.~(\ref{katharina}) with $f_1=e^-,\bar{f}_2=e^+$ (electrons and
positrons) and $B_1$ massless. Eq.~(\ref{ael}) is satisfied with 4
digits precision (more details in \cite{book}).

For the other zero, $a_1^{(3A)}$, one finds that it coincides,
with similar precision, with the electric coupling constant
\begin{equation}
a_1^{(3A)} =\alpha_{el}^d (3 m_d)=\frac{1}{9} \alpha_{el} (3 m_d),
\end{equation}
of $d$-quarks.

But what about $u$-quarks and neutrinos?
The interaction energy of the 3B string is plotted in Fig.~4.
\begin{figure}
\hspace{3cm}
\epsfig{file=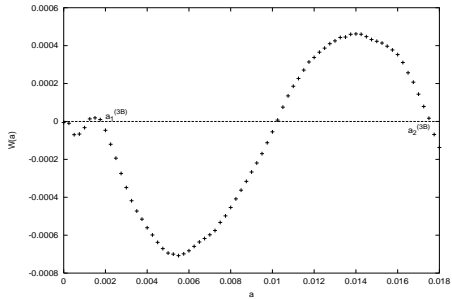, width=6cm, height=4cm} \caption{Same
as Fig.~3, but for the 3B string.}
\end{figure}
Again there are two zeros with negative slope in the low-coupling
region,
\begin{eqnarray*}
a_1^{(3B)} &=& 0.0018012(4) \\ a_2^{(3B)} &=& 0.017550(1)
\end{eqnarray*}
These, again with a precision of 4 digits, are found to coincide
with weak interaction strengths of $u$-quarks and
electron-neutrinos, if these are assumed to be there in addition
to electrically interacting $d$-quarks and electrons (see \cite{physicad,book}
for the details). At latest at this stage one notices that all
this can't be a random coincidence. One can have one random
coincidence, say of the fine structure constant with the zero
$a_2^{(3A)}$, but not 3 other random coincidences at the same
time! We are thus lead to the conclusion that the smallest zeros
of the interaction energy of the 3A and 3B string fix the
electroweak coupling strengths at the smallest fermionic mass
scales. 

Similarly, one numerically finds that the smallest zeros
of the interaction energies of the $N=2$ strings coincide with
strong couplings at the smallest bosonic mass scales. In
particular, the $W^\pm$ mass comes out correctly, and a Higgs mass
prediction of $(154.4\pm 0.5)$ GeV is obtained (see
\cite{physicad,book} for more details).

Another interesting observable is the self-energy $V^{(N)}(a)$ of
the strings. Typically the self-energies $V^{(N)}(a)$ have lots of
local minima. As an example, Fig.~5 shows $V^{(2)}(a)$ for the
2A/B string. For all strings, one numerically observes that
log-oscillatory behaviour with period $N^2$ sets in for small $a$,
hence e.g.\ for the $N=2$ strings all minima are only determined
up to an arbitrary power of 4. In other words, they are only
determined modulo 4. Remarkably, one observes minima that coincide
with Yukawa and gravitational couplings of the known fermions
modulo 4 (Fig.~5). The minima $b_2,b_6,b_{10}$ turn out to
coincide with Yukawa couplings modulo 4 of the heavy  fermions
$\tau ,b ,c$
\begin{equation}
b_i=\alpha_{Yu} =\frac{1}{4} \alpha_2 (m_H+2m_f) \left(
\frac{m_f}{m_W} \right)^2 \cdot 4^n,
\end{equation}
where $f=\tau , b, c$, respectively,\footnote{A $t$-quark minimum is also
observed, but outside the low-coupling region.}
\begin{figure}
\hspace{3cm}
\epsfig{file=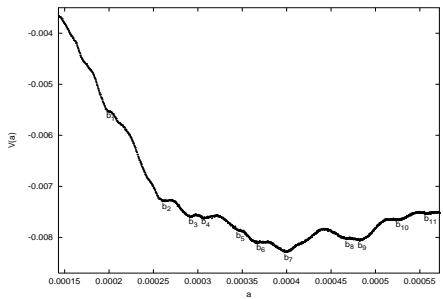, height=4cm, width=6cm} \caption{Local
minima of the self-energy of the 2A/B string fixing the fermion
masses.}
\end{figure}
and for the light fermions one observes that the self energy
has local minima for couplings that coincide with gravitational
couplings modulo 4. We numerically observe for $i=1,4,7,8,9$
\begin{equation}
b_i =\alpha_G = \frac{1}{2} \left( \frac{m_f}{m_{Pl}} \right)^2
\cdot 4^n,
\end{equation}
where $f=\mu,e,d,u,s$, respectively. Solving for $m_f$, one can
thus get fermion mass predictions modulo 2. The relevant power of
2 can then be obtained from other minima and additional symmetry
considerations \cite{book}. The remaining minima in Fig.~5 yield neutrino mass
predictions \cite{physicad,book}.

\section{Fixing standard model parameters}

What is the theory behind all these numerically observed
coincidences? The principal idea is very simple. At a very early
stage of the universe, where standard model parameters are not yet
fixed and ordinary space-time may not yet exist as well,
pre-standard model couplings are realized as coupling constants
$a$ in the chaotic noise space. The parameters are then fixed by
an evolution equation (a renormalization flow) of the form
\begin{equation}
\dot{a} = const \cdot W(a)  \label{franziska}
\end{equation}
(see Fig.~6), respectively
\begin{equation}
\dot{a}=-const \cdot \frac{\partial V}{\partial a} ,
\label{thomas}
\end{equation}
where we assume that the constant $const$ is positive. The
equations make {\em a priori} arbitrary standard model couplings
$a$ evolve to the stable zeros of $W(a)$, respectively to the
local minima of $V(a)$. There they will stay forever, since
any other value of the fundamental constants is energetically less
favourable.

\begin{figure}
\hspace{3cm}
\epsfig{file=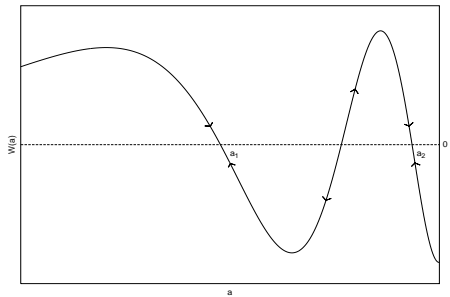, height=4cm, width=6cm}
\caption{Approach of arbitrary initial couplings to stable zeros
$a_1,a_2$ of the interaction energy (schematic plot).}
\end{figure}

Our main conclusion could be formulated as follows. The standard
model appears to have evolved to a state where its free parameters
minimize the vacuum energy associated with the chaotic noise
fields. If this chaotic dynamics keeps on evolving today, then the
fundamental constants are in fact stabilized by the local minima
of the energy landscape associated with the chaotic dynamics. Any
fluctuation to other values drives the fundamental 'constants'
immediately back to the equilibrium state, according to
eqs.~(\ref{franziska}) and (\ref{thomas}). The total expectation
of vacuum energy obtained from the chaotic dynamics in this way may well
correspond to the dark energy seen in the universe today.

\vfill\eject
\end{document}